\documentclass[
    ,final            % use final for the camera ready runs
  ]
  {aipproc}
\usepackage{epsfig}
\usepackage{amsmath}
\layoutstyle{6x9}

\begin{document}

\title{A Review of Nucleon Spin Calculations \\in Lattice QCD}


\author{Huey-Wen Lin}{
  address={Jefferson Laboratory, 12000 Jefferson Avenue, Newport
News, VA 23606}
}

\begin{abstract}
We review recent progress on lattice calculations of nucleon spin structure, including the parton distribution functions, form factors, generalization parton distributions, and recent developments in lattice techniques.
\end{abstract}

\maketitle

\section{Introduction}

Quantum chromodynamics (QCD) has been successful in describing many properties of the strong interaction. In the weak-coupling regime, we can rely on perturbation theory to work out the path integral which describes physical observables of interest. However, for long distances perturbative QCD no longer converges. Instead, we use a discretization of space and time in a finite volume to calculate these quantities from first principles numerically; such research forms the regime of lattice QCD\cite{DeGrand:2006zz}.

%\footnote{This paragraph is somewhat technical; interested readers can find more details in Ref.~\cite{DeGrand:2006zz} and references within.}% book
To keep the systematic error due to discretization under control, one follows Symanzik improvement order by order in terms of the ultraviolet cutoff ($a$) for both the action and operators.
However, the breaking of continuous (Euclidean) SO(4) symmetry allows many new degrees of freedom, leading to various lattice actions that return to the same continuum action once the symmetry is restored. Thus, there exist many gauge and fermion actions for us to choose from.
Today,
most gauge actions used are $O(a^2)$-improved and leave
small discretization effects ($O(a^3\Lambda_{\rm QCD}^3)$) due to gauge choices.
On the other hand, most fermion actions are only $O(a)$-improved and have systematic errors of $O(a^2\Lambda_{\rm QCD}^2)$ that become dominant. For this reason, lattice calculations are generally distinguished according to the fermion action used. Differences among the actions are benign once all systematics are included, and the choice of fermion action is constrained by limits of computational and human power and by the main physics focus. The commonly used actions are: domain-wall fermions (DWF), overlap fermions, Wilson/clover fermions, twisted-Wilson fermions and staggered fermions.

Since the real world is effectively continuous and infinitely large, we will have to take limits of $a \rightarrow 0$ and $V \rightarrow \infty$ to eliminate the artifacts introduced in a discretized finite box. With the most state-of-the-art supercomputer, we are close but yet to simulate at the physical pion mass. Using calculations at multiple heavier pion masses, which are affordable for available computational resources, we can apply chiral perturbation theory to extrapolate quantities of interest to the physical limit. A recent work by the BMW collaboration\cite{Durr:2008zz} calculating multiple lattice spacings, volumes and pion masses as light as 180~MeV provided an excellent demonstration of how ground-state hadron masses with fully understood and controlled systematics are consistent with experiment.
Such calculations with multiple pion masses also help to determine the low-energy constants of chiral effective theory.

A typical nucleon interpolating field used in the lattice calculation is
$
\chi_N = \sum_{\vec{x},a,b,c}
e^{i\vec{p}\cdot\vec{x}}\epsilon^{abc}
\left[u_a^T C\gamma_5 d_b \right] u_{c},
$
and the nucleon two- and three-point Green functions are obtained from
\begin{eqnarray}\label{eq:Greens}
\Gamma^{(2)}(t_{\rm src},t)&=& \langle \chi_N(t)\chi_N^\dagger(t_{\rm src})\rangle \\
\Gamma^{(3)}(t_{\rm src},t,t_{\rm snk}) &=& \langle \chi_N(t_{\rm snk},\vec p_{\rm snk})\,
{\cal O} (t,\vec{q})\,
\chi^\dagger_N(t_{\rm src},\vec p_{\rm src})\rangle,
\end{eqnarray}
where ${\cal O}$ is the operator of interest.
For the vector (axial) current, the operator is ${\cal O}=\overline{\psi}\gamma_\mu (\gamma_5) \psi$.
For the structure functions, the operators are
\begin{center}
\begin{tabular}{cccl}
$\langle x^n\rangle_{q}$: & ${\cal O}_{\mu_1...\mu_n}^q $&=&$i^{n-1}\overline{\psi}\gamma_{\{\mu_1}\overleftrightarrow{D}_{\mu_2}\cdots\overleftrightarrow{D}_{\mu_n\}}\psi$\\
$\langle x^n\rangle_{\Delta q}$: & ${\cal O}_{\mu_1...\mu_n}^{5q}$&=&$i^{n-1} \overline{\psi} \gamma_{\{\mu_1} \gamma_5 \overleftrightarrow{D}_{\mu_2}\cdots \overleftrightarrow{D}_{\mu_n\}}\psi$\\
$\langle x^n\rangle_{\delta q}$: & ${\cal O}_{\mu\mu_1...\mu_n}^{\sigma q}$&=&$i^{n-1}
\overline{\psi} \gamma_5\sigma_{\mu \{\mu_1}\overleftrightarrow{D}_{\mu_2}\cdots \overleftrightarrow{D}_{\mu_n\}}\psi$,\\
\end{tabular}
\end{center}
where $\overleftrightarrow{D}=\frac{1}{2}(\overrightarrow{D}-\overleftarrow{D})$ is the difference between forward and backward covariant derivatives.
We calculate only the ``connected'' diagrams, which means the inserted quark current is contracted with the valence quarks in the baryon interpolating fields, as in the majority of lattice three-point calculations. However, due to isospin symmetry, isovector quantities have a cancellation that removes the unknown disconnected piece. 
%\footnote
(Disconnected diagrams are notoriously difficult to calculate directly on the lattice.
They require that expensive fermion-matrix inversion be applied to numerous source vectors. Much effort has been devoted to solving this difficulty in the near future with new techniques.)
For more details, please refer to a selection of plenary talks: Refs.~\cite{plenary} and references within.

\section{Lattice QCD Reveals the Structure of the Nucleon}

\begin{figure}
\includegraphics[height=.21\textheight]{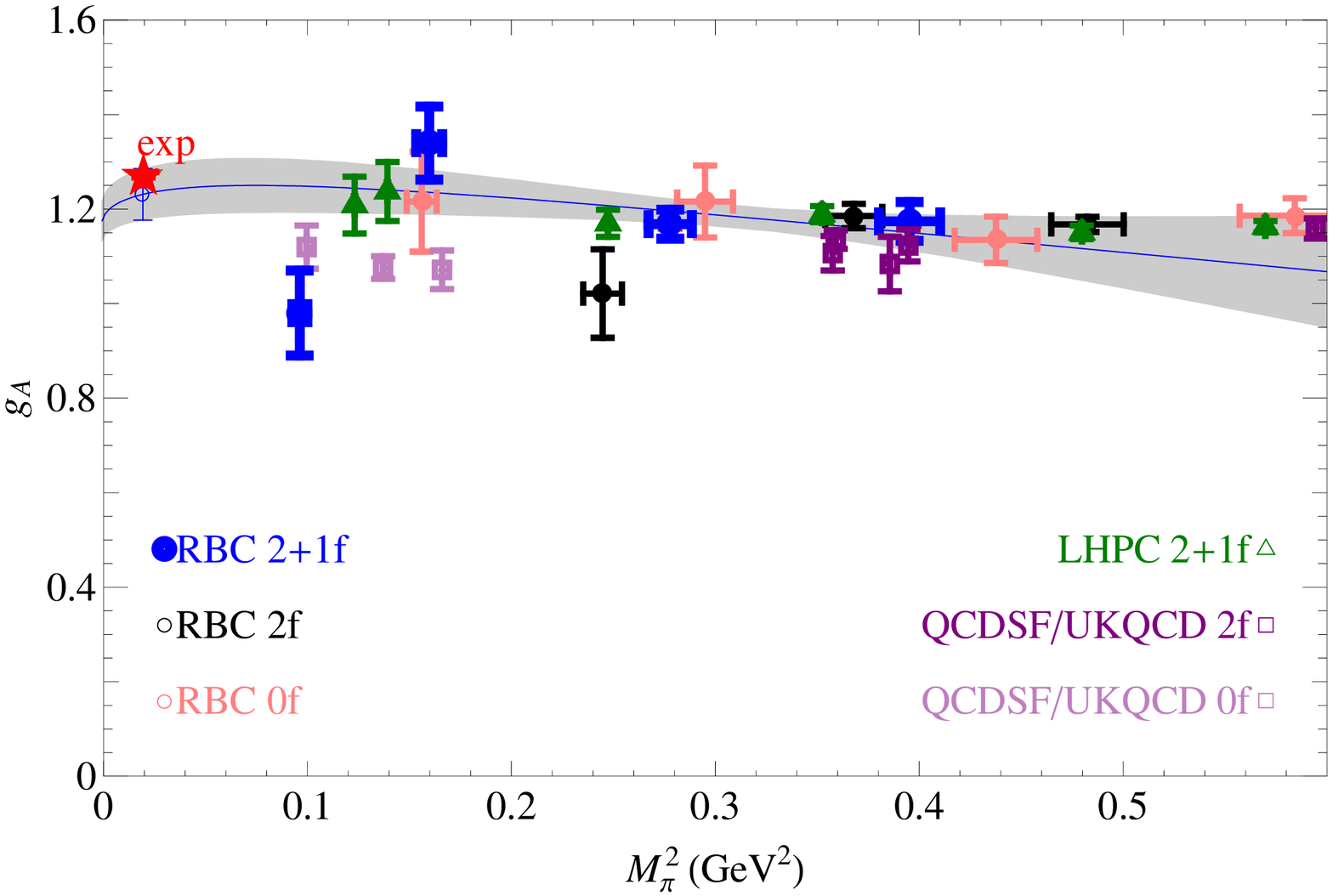}
\includegraphics[height=.21\textheight]{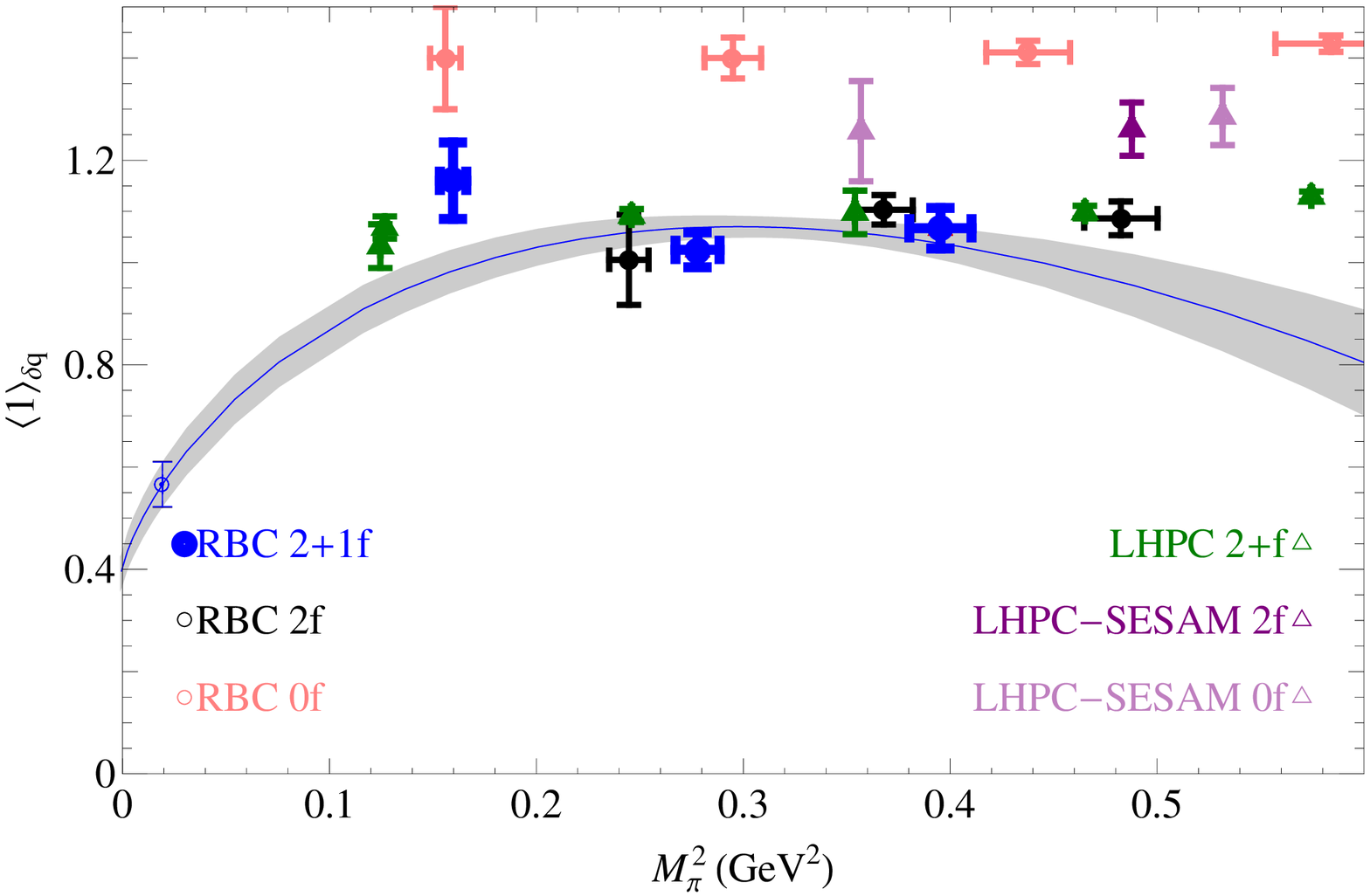}
\caption{(left) Renormalized axial charge versus
pseudoscalar mass squared from various lattice groups. 
(right) Zeroth moment of transversity from different lattice groups. The bands in both plots are chiral extrapolations fit to the RBC 2+1f data.
 } \label{fig:gA_all}
\end{figure}

The nucleon axial charge is well measured in neutron $\beta$-decay experiments, so it is a natural candidate for demonstrating how well lattice QCD can be extrapolated to the physical pion point. The isovector axial charges $g_A$ are defined as the zero-momentum-transfer limits of the
isovector axial form factors. Results from various collaborations are shown on the left-hand side of Figure~\ref{fig:gA_all}. We show a small-scale--expansion fit (gray band) to the 2+1-flavor (DWF sea quarks) RBC data\cite{RBC} (blue filled circles). The results are consistent with the LHPC data\cite{LHPC} (DWF valence with staggered sea); RBC's 2f and 0f results are consistent with QCDSF's 2f Clover and 0f overlap fermion numbers\cite{QCDSF} respectively. The lowest--pion-mass values among RBC's 2+1f and 2f results may suffer from sizable finite-volume systematic errors; larger-volume calculations should be carried out to confirm these suspicions.

Results for the zeroth moment of transversity $\langle 1 \rangle_{\delta q}$ are given on the right-hand side of Figure~\ref{fig:gA_all}. We observe rather weak dependence (roughly linear) on the quark mass, which remains consistent for calculations with the same number of sea-quark flavors.
The chiral extrapolation form\cite{xChPT} %{Detmold:2001jb,Arndt:2001ye,Chen:2001gr,Detmold:2002nf}
is applied to the RBC 2+1f data\cite{RBC}, yielding 0.56(4) at the physical pion mass.
However, this is close to what has been found by LHPC with mixed action\cite{LHPC}: about 0.7.
These extrapolated values are significantly smaller than those found at the simulated pion masses, which are near 1. We urgently need data at the lightest pion mass to confirm the rapidly decreasing behavior predicted by the chiral effective theory.

\begin{figure}
\includegraphics[height=.21\textheight]{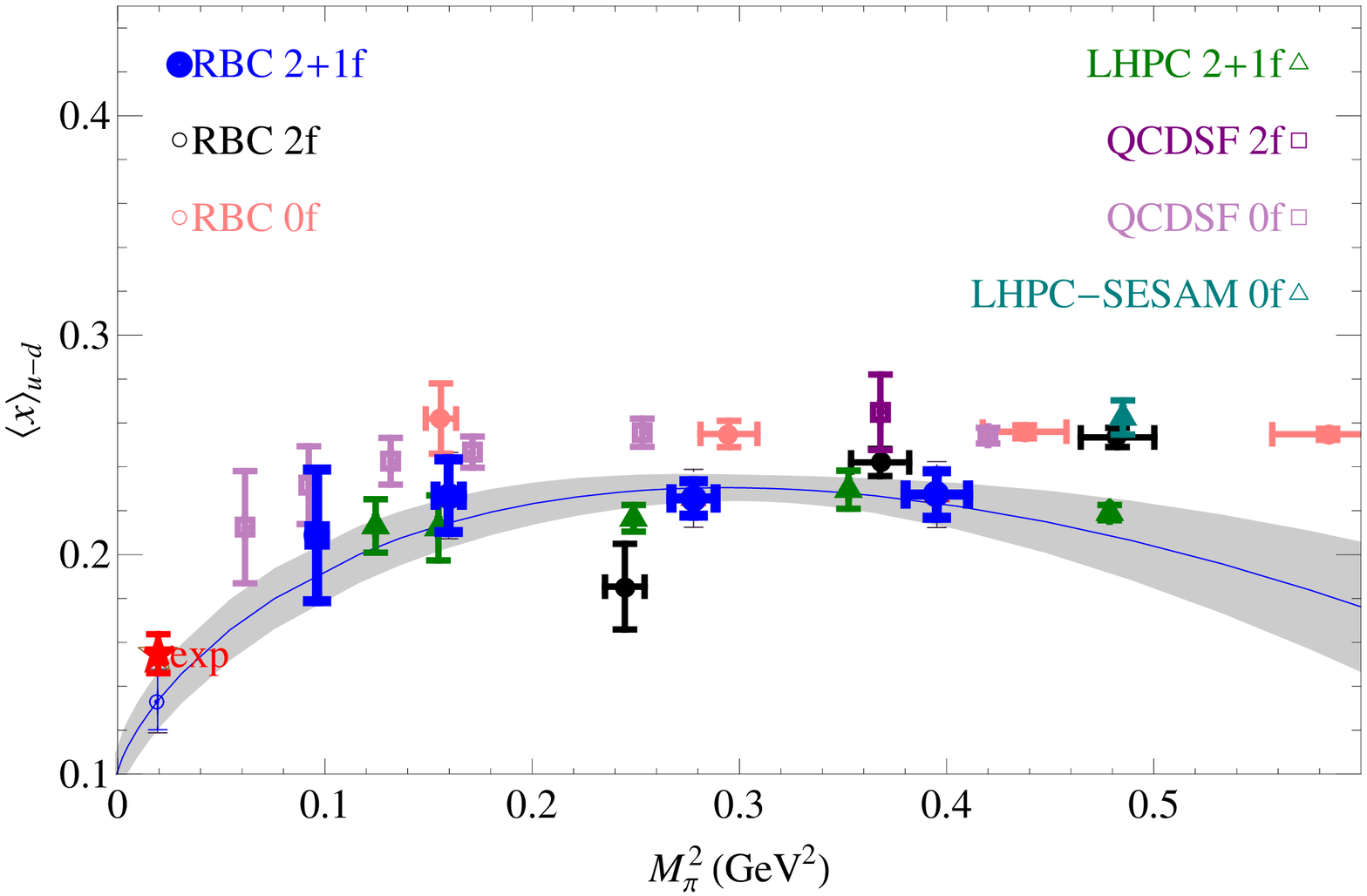}
\includegraphics[height=.21\textheight]{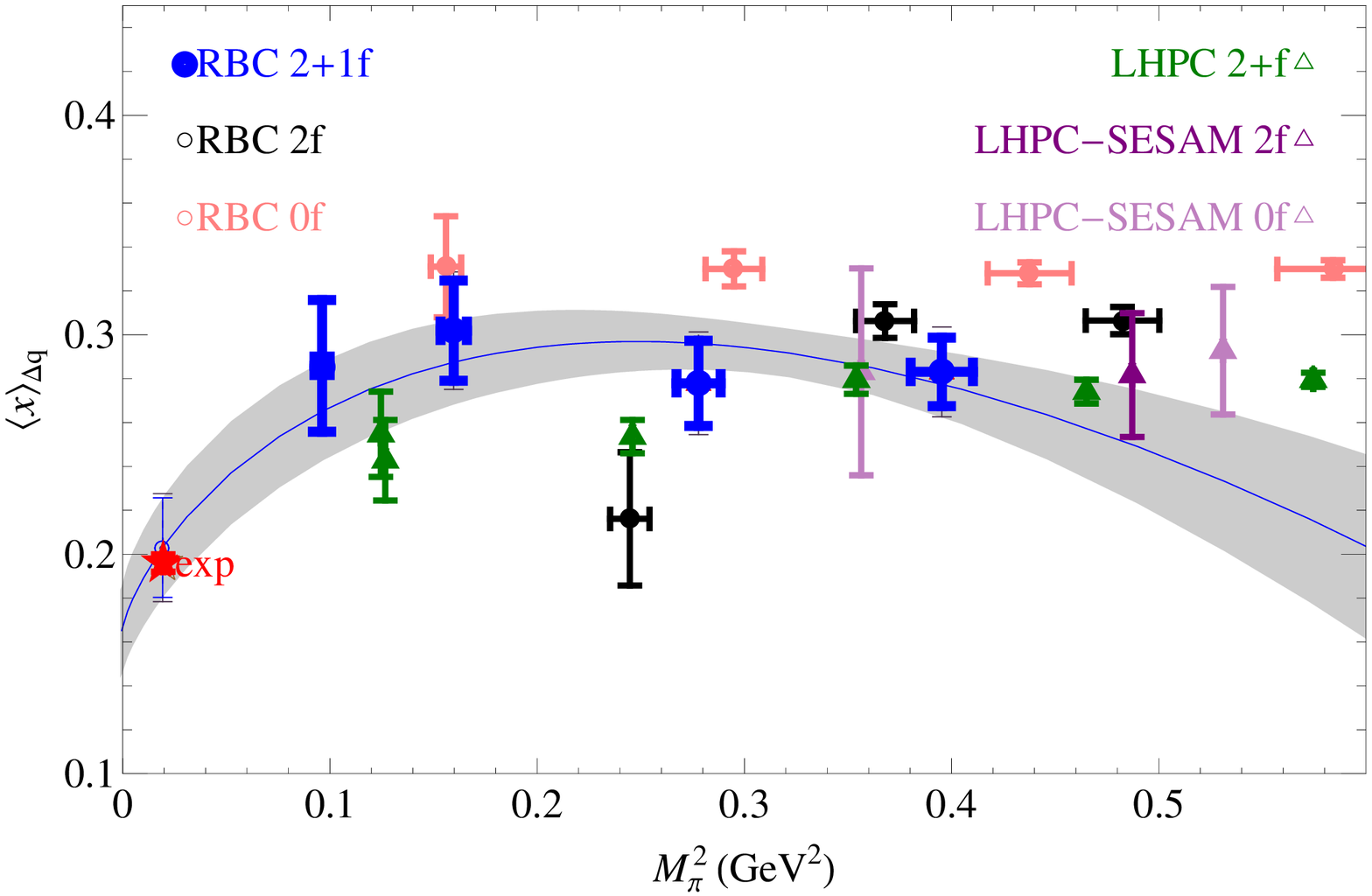}
\vspace{-0cm} \caption{Global comparison of the first moments of the quark momentum fraction (left) and helicity (right) distributions in terms of $M_\pi^2$ and their chiral extrapolations. The bands in both plots are chiral extrapolations fit to the RBC 2+1f data.} \label{fig:allMoments}
\end{figure}

Figure~\ref{fig:allMoments} shows the latest calculations of the first moments of the quark momentum fraction (left) and helicity (right) distributions. Here again the 0f, 2f and 2+1f results are compared between different collaborations with different choices of fermion actions and are seen to be consistent among calculations with the same number of sea-quark flavors.
The chiral extrapolation\cite{xChPT} is performed using RBC's 2+1f data, which gives 0.133(13) and 0.203(23) for the first moments of the quark momentum fraction and helicity distributions respectively, consistent with experiment.
We see strong curvature due to the chiral form; more light-pion points should be taken to reduce extrapolation uncertainties.
These extrapolation numbers are also consistent with the LHPC's mixed-action calculation\cite{LHPC}.

Higher moments of the isovector distributions are calculated by LHPC (SCRI, SESAM) with  0f and 2f Wilson and clover fermions\cite{LHPC} %{Dolgov:2002zm}
and by QCDSF with 0f clover fermions at multiple lattice spacings\cite{QCDSF}. %{Gockeler:2004wp}.
Consistent results are seen among different groups:
The second and third moments are about 25\% and 10\% of the first moments respectively.
However, for moments with $n\geq4$, divergences occur involving lower-dimension operators at finite $a$, which limits the number of moments accessible to lattice QCD.

The twist-3 first moment of the polarized structure function $d_n$ is another interesting feature to consider. It is related to the polarized structure functions $g_1$ and $g_2$ and the Wandzura-Wilczek relation\cite{Wandzura:1977qf}.
The lowest moment $d_1$ from RBC's 2+1f data extrapolated to the physical pion mass is consistent with zero $d_1^{\rm bare}=-0.002(2)$.
Combined with the small value of $d_2$ found by QCDSF's 2f calculation\cite{QCDSF},  %{Gockeler:2000ja},
we conclude that the Wandzura-Wilczek relation between the moments of $g_1$ and $g_2$, which asserts vanishing $d_n$, is at least approximately true.

\begin{figure}
\includegraphics[height=.34\textheight]{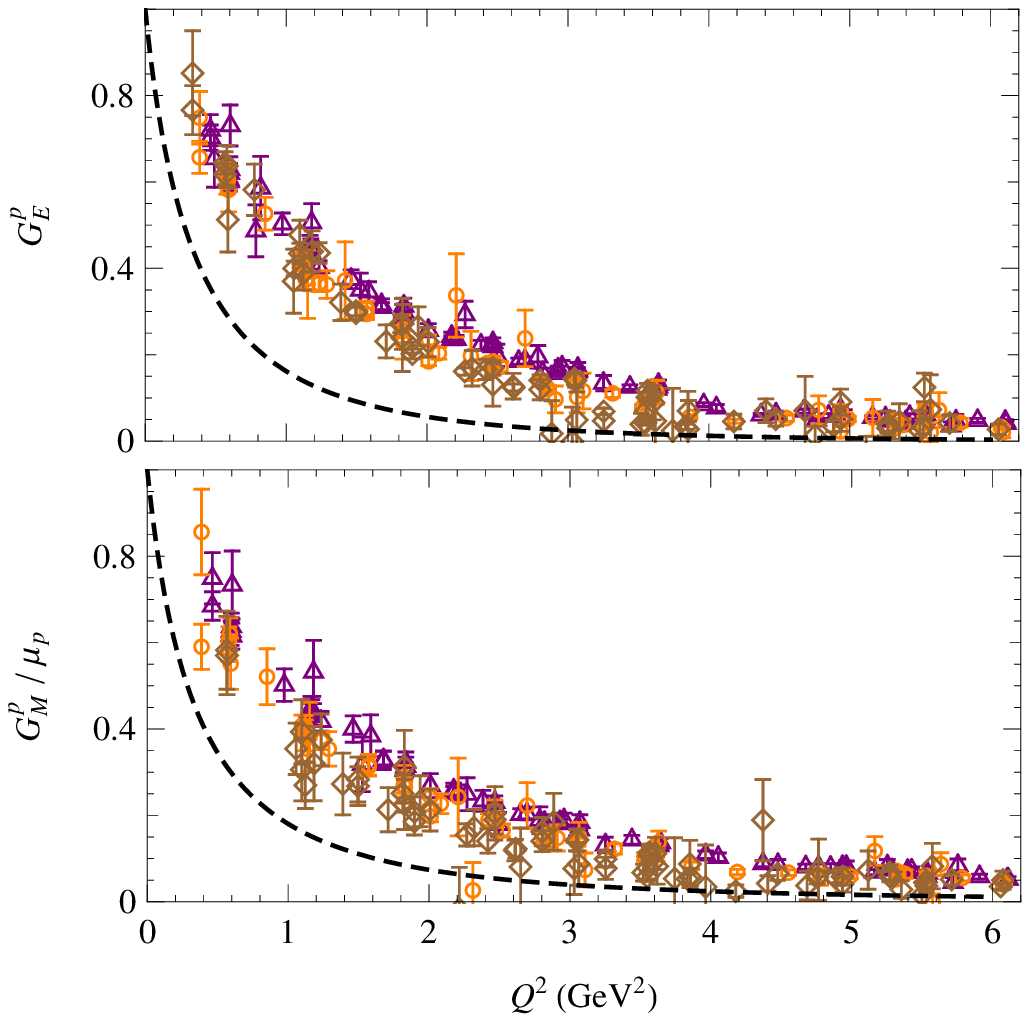}
\includegraphics[height=.34\textheight]{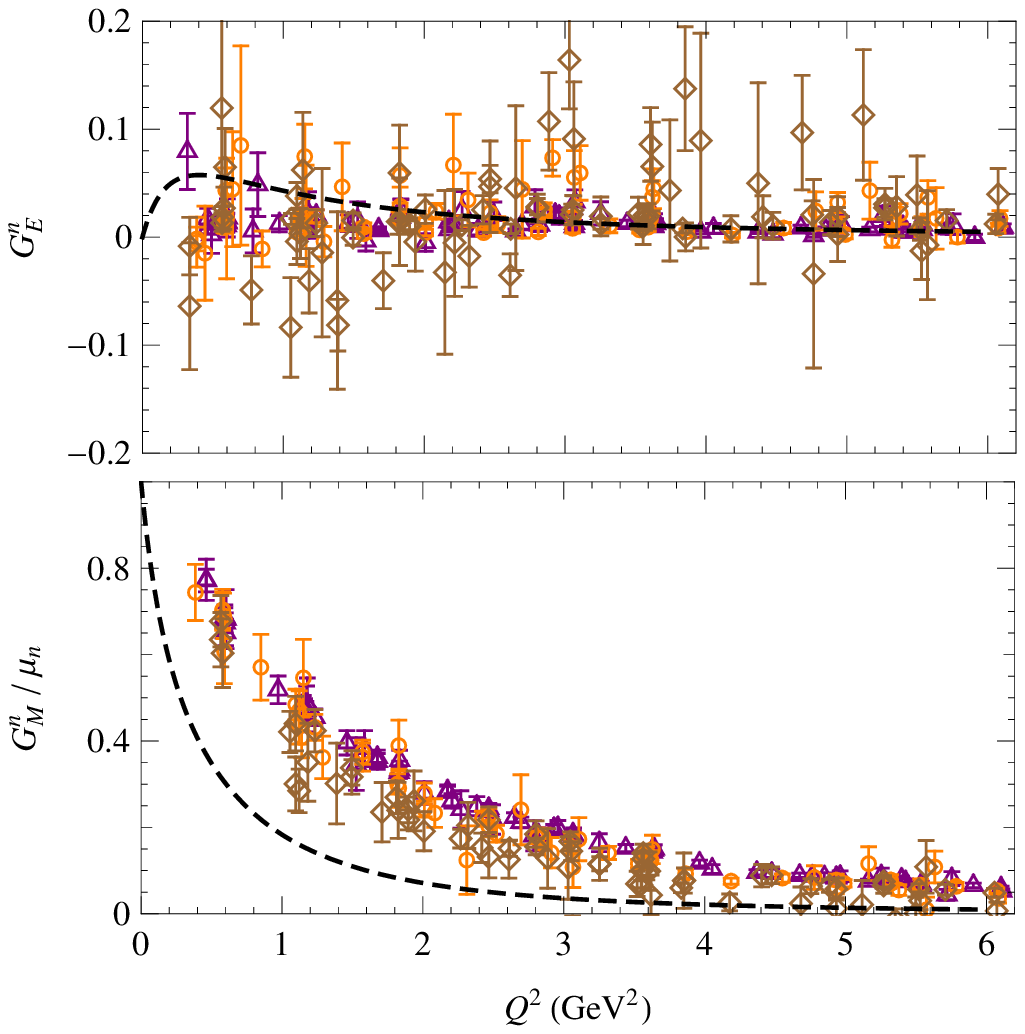}
\vspace{-0 cm} \caption{Nucleon form factors with pion masses of 480 (triangles), 720 (circles) and 1080 (diamonds) MeV. The dashed lines are plotted using experimental form-factor fit parameters\cite{FFs}.%{Arrington:2007ux,Kelly:2004hm}
} \label{fig:FF}
\end{figure}

Studying the momentum-transfer ($Q^2$) dependence of the elastic electromagnetic form factors is important in understanding the structure of hadrons at different scales. There have been many experimental studies of these form factors on the nucleon. A recent such experiment, the Jefferson Lab double-polarization experiment (with both a polarized target and longitudinally polarized beam) revealed a non-trivial momentum dependence for the ratio $G_E^p/G_M^p$. This contradicts results from the Rosenbluth separation method, which suggested $\mu_p G_E^p/G_M^p \approx 1$. The contradiction has been attributed to systematic errors due to two-photon exchange that contaminate the Rosenbluth separation method more than the double-polarization.  (For details and further references, see the recent review articles: Refs.~\cite{FFs}%{Arrington:2007ux,Perdrisat:2006hj,Arrington:2006zm}
.)
Lattice calculations can make valuable contributions to the study of nucleon form factors, since they allow access to both the pion-mass and momentum dependence of such form factors.
Recently, the limitations of the largest-available $Q^2$ (in terms of the quality of the signal-to-noise ratios) has been overcome\cite{Lins}%{Lin:2008qv,Lin:2008gv}
. An exploratory study using clover fermions extends the range of momentum transfer to 6~GeV$^2$, as shown in Figure~\ref{fig:FF}. The $Q^2$ dependence of the neutron has exceeded the range of the current existing data. Such calculations will provide interesting comparisons for data collected after the future 12-GeV upgrade at Jefferson Lab.

%%%%%%%%%%%%%%%%%%%%%%%%%%%%%%%%%%%%
% GPD's
%%%%%%%%%%%%%%%%%%%%%%%%%%%%%%%%%%%%%%

\begin{figure}
\includegraphics[height=.26\textheight]{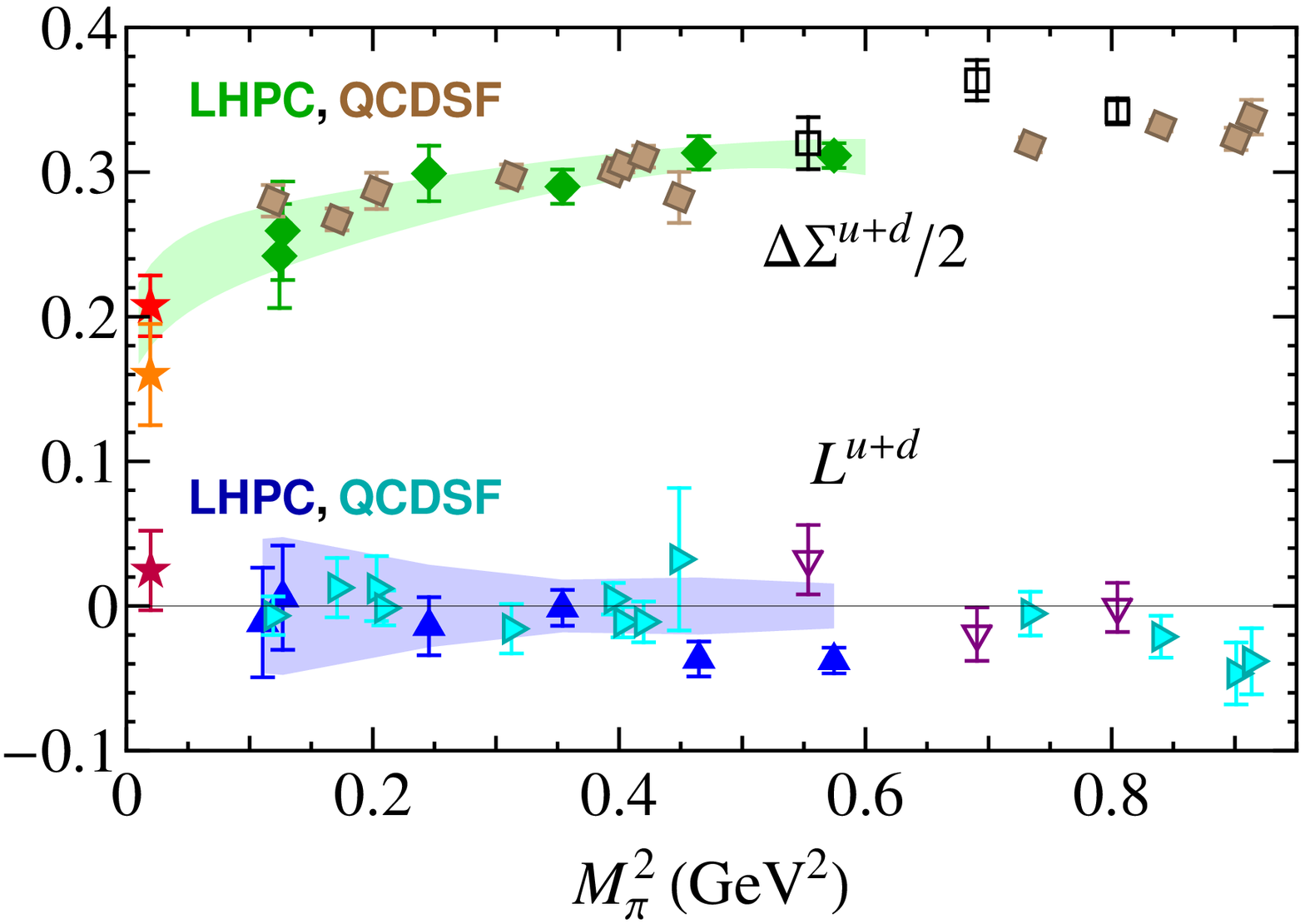}
\includegraphics[height=.26\textheight]{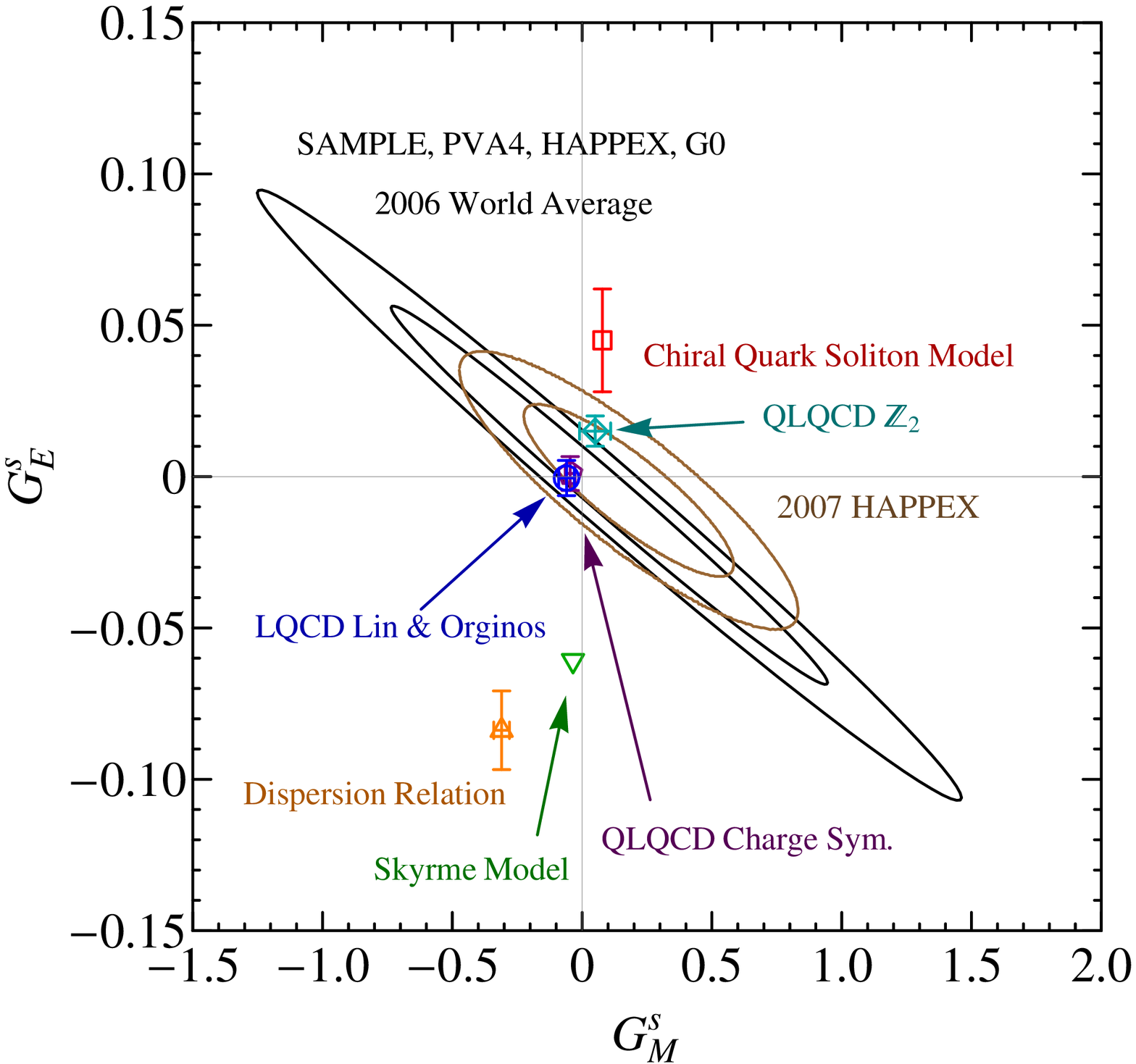}
\vspace{-0cm} \caption{(left) The quark-component contributions to quark spin and orbital angular momentum in the spin of the nucleon from LHPC (squares, diamonds and upward triangles) and QCDSF (tilted squares and rightward triangles).
The filled/open symbols represent dynamical/quenched data. The bands are chiral extrapolations based on LPHC's dynamical data. 
%QCDSF 
(right) Experimental results (ellipses) and various theoretical calculations (points) of $G_M^s$-$G_E^s$} \label{fig:Spin-GEMs}.
\end{figure}

The generalized parton distributions (GPDs) have been calculated by LHPC (2+1f mixed action, $M_\pi \sim 350$--760~MeV)\cite{LHPC} %{Hagler:2007xi}
and QCDSF (2f clover action, $M_\pi \sim 340$--950~MeV)\cite{QCDSF}. %{Ohtani:2007}.
One of the main topics of physics interest derived from GPDs is to study the origins of the nucleon's spin. The quark spin ($\Delta \Sigma^q$) and total quark contribution ($J^q$) to the angular momentum can be connected to GPDs via\cite{Ji:1996ek}
\begin{equation}
\frac{1}{2}\Delta \Sigma^q = \frac{1}{2}\tilde{A}^q_{10}(0); \;\;
J^q=\frac{1}{2}[{A}^q_{20}(0)+{B}^q_{20}(0)], %-\frac{1}{2}\Delta \Sigma^q \label{ionflux}
\end{equation}
where $\tilde{A}^q_{n0}$ and $\{A,B\}^q_{n0}$ are the polarized and unpolarized generalized form factors.
The left-hand side of Figure~\ref{fig:Spin-GEMs} shows the quark spin and orbital angular momentum ($L^q$) calculated by LHPC and QCDSF; they are consistent.
One found the total $d$ quark angular momentum and the angular momentum from the sum of $u$ and $d$ quarks to be consistent with zero. This is because $L^d$ and $L^u$ are of the same magnitude but have opposite signs. A similar relation holds for $\Delta \Sigma^d$ and $L^d$.
Note that in both of the calculations only connected diagrams are included.

For the transverse structure of the proton, QCDSF calculates the lowest two moments of the transverse spin densities of quarks in the nucleon\cite{QCDSF}%{Gockeler:2006zu}
. An exploratory attempt has been made to calculate transverse-momentum distributions (TMDs) using lattice link products to approximate the Wilson line in light-cone frame in spatial coordinates and Fourier transforming into momentum-space to map the density distributions\cite{LHPC}; %{Musch:2008jd};
this calculation used 2+1f mixed action (DWF on staggered, $M_\pi \sim 500$~MeV).

Another new development during the last year is progress on disconnected diagrams. An indirect way of calculating the strangeness form factors via charge symmetry in 2+1f lattices is shown on the right-hand side of Figure~\ref{fig:Spin-GEMs}. The direct approach has also made great progress. For example, the strange-quark distribution $\langle x \rangle_s$ has been calculated by the $\chi$QCD collaboration on 2+1f lattices\cite{chiQCD}%{Deka:2008xr,Doi:2008hp}
. The same group has also calculated the gluon momentum fraction $\langle x \rangle_g$ with improved signal.

This is an exciting era for the use of lattice QCD in nuclear physics: there have been huge leaps due to increasing computational resources worldwide and improved algorithms, allowing continual improvement in lighter pion masses, larger volumes and finer lattice spacings. Various groups have demonstrated universality with the consistent results coming from independent calculations using different lattice actions. By reproducing well measured experimental values, we solidify confidence in lattice predictions of quantities that have not or cannot be measured by experiment. There are many different aspects of hadron structure that one can do with lattice QCD; only a few examples have been presented here.

In the near future, pion masses around 200~MeV or lighter with multiple volumes and lattice spacings will become commonplace. There will be less need to depend on chiral perturbation theory for extrapolations, and once the physical pion mass becomes accessible, we can check its correctness.
Full-contraction calculations (including disconnected diagrams) in all matrix elements, form factors and GPDs will lead to precision calculations for individual quark components or individual proton and neutron quantities. Further improvements in methodology and expanding computational resources will allow direct calculations of the gluon content of the nucleon. The future is unlimited with lattice QCD.

{{\it Acknowledgements}: Authored by Jefferson Science Associates, LLC under U.S. DOE Contract No. DE-AC05-06OR23177. The U.S. Government retains a non-exclusive, paid-up, irrevocable, world-wide license to publish or reproduce this manuscript for U.S. Government purposes.}

%%%%%%%%%%%%%%%%%%%%%%%%%%%%%%%%%%%%%%%%%%%%%%%%%%%%%%%%%%%%%%%%%%%%%%%%%%%%%%%%%%%%
\bibliographystyle{aipproc}
{\vspace{-0.5cm}}
\hyphenation{Post-Script Sprin-ger}

\end{document}